# The Andromeda Galaxy's most important merger ~ 2 Gyrs ago as M32's likely progenitor

Richard D'Souza* & Eric F. Bell

**Although the Andromeda Galaxy's (M31) proximity offers a singular opportunity to understand how mergers affect galaxies[1], uncertainty remains about M31's most important mergers. Previous studies focused individually on the giant stellar stream[2] or the impact of M32 on M31's disk[3,4], thereby suggesting many significant satellite interactions[5]. Yet, models of M31's disk heating[6] and the similarity between the stellar populations of different tidal substructures in M31's outskirts[7] both suggested a single large merger. M31's outer low-surface brightness regions (its stellar halo) is built up from the tidal debris of satellites[5] and provides decisive guidance about its important mergers[8]. Here we use cosmological models of galaxy formation[9,10] to show that M31's massive[11] and metal-rich[12] stellar halo, containing intermediate-age stars[7], dramatically narrows the range of allowed interactions, requiring a single dominant merger with a large galaxy ($M_*$~$2.5\times10^{10}$ $M_\odot$, the third largest member of the Local Group) ~2 Gyr ago. This single event explains many observations that were previously considered separately: its compact and metal-rich satellite M32[13] is likely to be the stripped core of the disrupted galaxy, M31's rotating inner stellar halo[14] contains most of the merger debris, and the giant stellar stream[15] is likely to have been thrown out during the merger. This interaction may explain M31's global burst of star formation ~2 Gyr ago[16] in which ~1/5 of its stars were formed. Moreover, M31's disk and bulge were already in place suggesting that mergers of this magnitude need not dramatically affect galaxy structure.**

In order to determine what M31's massive and metal-rich stellar halo (Fig. 1) tells us about its most important merger, we use models of galaxy formation encoding a diversity of merger (or accretion) histories from two independent set of simulations[9,10]. Simulated galaxies with M31's stellar and virial mass are predicted to have a wide range of accreted masses (sum of the stellar masses contributed by all disrupted satellites) and consequently a wide range of masses, metallicities and accretion times of the most massive progenitors (Fig. 2a,b). We use M31's outer stellar halo (R > 27 kpc, which is not contaminated by kicked-up disk stars[7]) to infer the mass and metallicity of M31's total accreted stellar component ($M_{*,acc}$ > 2 x $10^{10}$ $M_\odot$, $[M/H]_{acc}$ > -0.3; see

Methods). We isolate a subsample of true 'M31 analogues' from the models by requiring a large accreted mass similar to M31's. By virtue of their large accreted masses, these M31 analogues have a much narrower range of accretion histories, invariably having a *massive* ($10^{10} < M_*/M_\odot < 10^{11}$) and *metal-rich* (-0.2<[M/H]<0.2) progenitor accreted in the last ~5 Gyr. These massive progenitors are in general star-forming, gas-rich, rotating galaxies with well-established metallicity gradients. Moreover, the high metallicity of M31's accreted stellar component argues strongly that a single massive progenitor dominates M31's accreted mass budget (Fig. 2c).

We use M31 analogues from a hydrodynamical simulation[9] to illustrate the merger process and identify likely debris, in the hope of further constraining the time of merger and the type of progenitor. As the merger progresses, the satellite galaxy is disrupted while experiencing a burst of star formation (illustrative example in Fig. 3; see Methods). The cessation in centrally-concentrated star formation occurs shortly before coalescence with the main galaxy. Most of the satellite is disrupted into a structured, but highly-flattened rotating inner stellar halo with an exponential density profile along the major axis, a $R^{1/4}$ profile along the minor axis, and a velocity dispersion that drops off towards larger radii. The debris field is metal-rich. Gradients in the progenitor result in the stellar halo having variations in metallicity by a factor of 10 and time of star formation shut-off by > 2 Gyr, with the most metal-rich and youngest parts concentrated towards the center. In most cases, a prominent metal-rich tidal stream is produced.

The observational features most consistent with the metal-rich debris of M31's massive accreted progenitor are M32, the inner stellar halo and the giant stellar stream (Fig. 1), while excluding the other metal-poor satellites and streams that are expected to be from numerous smaller accretion events[5]. We hypothesize that all these three metal-rich features stem from M31's massive accreted progenitor. We now discuss how these features can be used to further constrain the type of progenitor and to approximately time the merger.

**M32:** M32's compact size, solar metallicity, prolonged star formation history and sizeable intermediate-age stellar populations[13] are easily explained if it is the compact core of M31's massive (~2.5 x $10^{10}$ $M_\odot$) gas- and metal-rich accreted progenitor (Fig. 4) which we term as **M32p**. M32p is likely to have experienced a burst of late star formation triggered by gas inflow to the center of the galaxy as it is being accreted (Fig. 3a, top right), similar to the observed star formation experienced by M32 ~2-5 Gyr ago (Fig. 1)[13]. M32's relatively low-mass central black hole ($3.4 \pm 1.6 \times 10^6$ M$_\odot$) [17] implies that M32p originally had a likely low-mass bulge ($M_{Bulge} < 5.9 \times 10^9$ M$_\odot$), using the black-hole mass-bulge mass relationship[18]. Such a small bulge would have experienced little dynamical friction, consistent with M32 not sinking to the center of M31. If M32 is indeed the core of the massive accreted progenitor of M31, then the timing of the starburst allows us to approximately date the interaction: M32p possibly started interacting with M31 approximately 5 Gyr ago and its disruption continued until around 2 Gyr ago.

**Inner stellar halo**: M32p's tidally-stripped centrally-concentrated debris shares many of the properties of M31's metal-rich inner stellar halo, including its flattened spheroidal nature, its disk-like rotation[14], the presence of intermediate-age stars and some of its stellar population variations[7] (Fig. 3). While stellar population studies suggest that the inner halo is a mixture of disk and accreted material[7], M32p's debris dominates the minor axis density profile of M31 from a projected distance of ~8 kpc out to 25 kpc [19,20], suggesting that it forms a major component of the inner stellar halo in this radial range (Fig. 3b).

**Giant stellar stream:** Since large metal-rich tidal streams are frequently produced by the most-massive satellite (Fig. 3; see Methods), we suggest that it is likely that the giant stream is from M32p's tidal disruption. The similarity in stellar populations between the giant stream and the inner stellar halo[7] as well as the synchronous star formation burst ~2 Gyrs ago evident in the high-metallicity accreted populations of the giant stream, the inner stellar halo and M32 (compare Figs. 1 & 3) increases dramatically the probability of this association. If this is so, then the giant stream offers the potential to constrain the orbit of M32p.

If the giant stream stems from M32p, then M32's position and line-of-sight velocity should be consistent with the forward orbit of the stream's progenitor. Uncertainties in distances and the nature of the stream's progenitor[2] make it challenging to model the progenitor's orbit (see Supplement Section 4), and a progenitor has not been found. Despite these uncertainties, there is strong evidence that the North-Eastern and Western shelves trace out the stream's forward orbit[2]. M32 is consistent with the positional and radial velocity constraints of these shelves[2], and thus could be the core of the giant stream's progenitor. This proposition can be tested using future measures of M32's proper motion, coupled with future models of the stream that incorporate suitable priors on the progenitor's mass, rotation and central density.

Our inferences about M32p, the third largest member of the Local Group (Fig. 4), give further weight to recent attempts to use the steep age-velocity dispersion relationship of M31's disk to constrain the accretion of a recent (1.8-3 Gyr) massive progenitor (1:4) [6]. These major-merger simulations also reproduce the general properties of the giant stream. However, these simulations assume that M31's bulge and bar were also created by this merger, and consequently the nuclei of the two galaxies are forced to coalesce. Not only have we enriched this picture by suggesting that the core of that massive accreted progenitor survived to the present day as the metal-rich compact M32, and consequently did not form M31's bulge and bar, but we have also provided direct unambiguous evidence of this recent major merger from M31's stellar halo.

Our formation scenario for M32 impacts the debate about its origin. Compact elliptical galaxies (~$10^9$ $M_\odot$; Supplementary Section 3) are believed to be the stripped cores of previously more massive galaxies[21]. Yet, it has also been suggested that compact ellipticals might have formed in a starburst followed by a violent collapse, with no stripping involved[22]. We suggest that the evidence proposed for an intrinsic origin of M32 is ambiguous. M32's absence of tidal features[23] can be easily explained if it is the compact dense core of M32p allowing it to resist further stripping. Built up from the small bulge of its progenitor, M32 naturally obeys the structural scaling relations of classical bulges and ellipticals[24]. Evidence of tangential anisotropy in M32's outer velocity dispersion[25] is consistent with the preferential tidal stripping of its stars[26]. While M31-M32 interaction models designed to reproduce M31's long-lived 10 kpc star-forming ring have been used to support M32's intrinsic origins[3,4], major merger simulations can also reproduce M31's star-forming ring[6]. In contrast, M32's unique properties make it stand out among other compact ellipticals and argues strongly for a stripped origin for at least M32. M32 has the smallest size among all known compact ellipticals. Not only does its extended star formation history and starburst 2-4 Gyr ago[13] rule out an intrinsic formation at a higher redshift, but these properties are also predicted by models that tidally strip gas-rich progenitors with compact cores[21]. M32 has a very high metallicity for its stellar mass suggesting it was once much more massive. Our work advances this debate in two primary ways. First, we suggest that M31's stellar halo is the reservoir of much of M32's stripped material and provides decisive guidance to constrain its formation history. Second, by identifying M32's progenitor, we find that the rarity of M32-like objects in the local Universe is set by the number density of M32p analogues convolved with their merger rate since z~2 (see Methods).

In addition to laying out a framework for characterizing the most massive merger events in other galaxies, there are several interesting implications of this work for our understanding of M31. First, because M31's disk pre-dates this interaction[16], M31's disk survived a merger with mass ratio between ~0.1 and ~0.3 [6]. Second, as demonstrated by recent simulations, this major merger may be responsible for the thickening of M31's disk to its present scale height of ~1 kpc [27] as well as the steepness of its stellar age velocity-dispersion relationship[6]. Third, the timing of the merger suggests that it caused a galaxy-wide star formation episode in M31's disk ~2 Gyr ago in which ~1/5 of its stars formed[16]. If indeed this episode is associated with its merger with M32p, this provides the first empirical measurement of the lifecycle effects of such a merger. Finally, large bulges like M31's have been suggested to have been made in galaxy mergers[1]. Yet, M31 had already formed its bulge stars >6 Gyr ago[28], long before M31's merger with M32p. This adds to the evidence that merging and bulge formation are not trivially linked.

**Acknowledgments**: We would like to thank the reviewers, Julianne Dalcanton, Andrew Cooper, Antonela Monachesi, Scott Trager, Monica Valluri, Kohei Hattori, Kathryn Johnston, Puragra Guhathakurta, Brian Devour, Simona Vegetti, Jessie Runnoe, Megan Reiter, Paul Mueller, Joe Wagner and Robert Macke for comments on the draft. We thank Andrew Cooper for access to his particle-tagging simulations. We thank Rodrigo Ibata and Karoline Gilbert for permission to use their figures in this publication.

**Author Information**

**Affiliations**

*Department of Astronomy, University of Michigan, 311 West Hall, 1085 S. University, Ann Arbor, MI 48109-1107, USA*

Richard D'Souza and Eric Bell

*Vatican Observatory, Specola Vaticana V-00120, Vatican City State*

Richard D'Souza

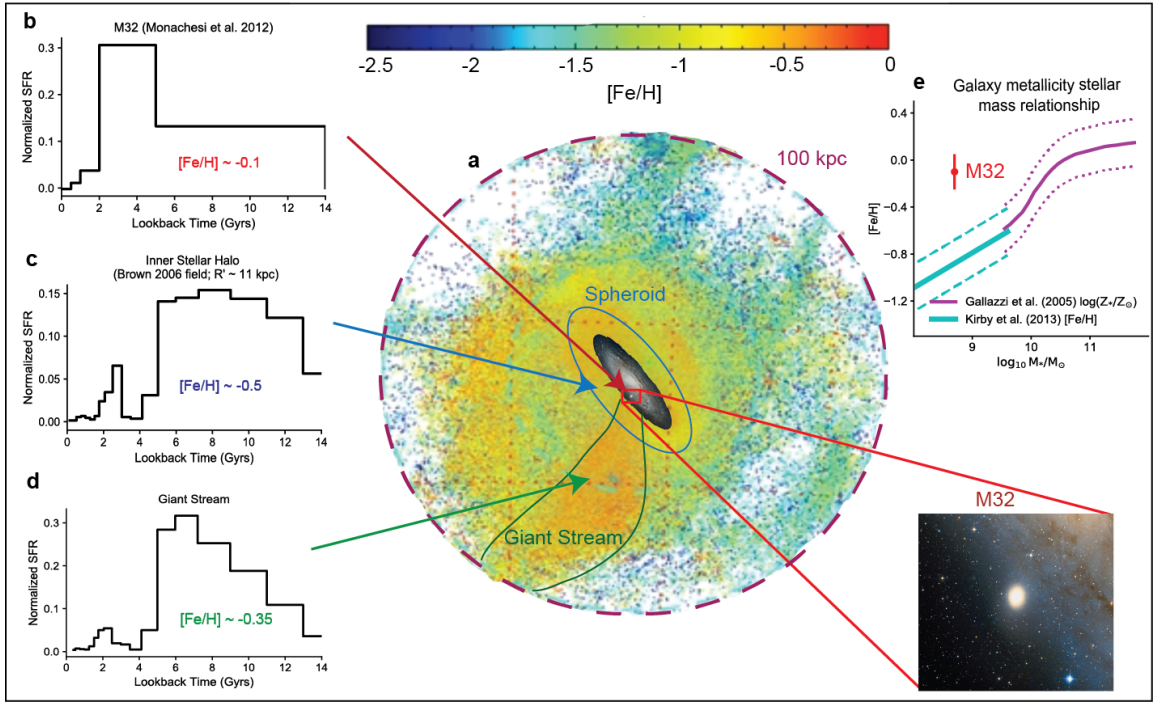

**Figure 1: The massive and extensive metal-rich stellar halo of M31 contains a substantial population of young and intermediate-aged stars.** a) The metallicity-map of the stellar halo of M31 extending out to 100 kpc is adopted from the PAndAS survey[11]. The intensity scale of the metallicity map codes the density of stars and the color denotes the typical median metallicity of the halo stars. M31's stellar halo is 10x more metal-rich and 20x more massive than the Milky Way's. In particular, three metal-rich 'accreted' features are highlighted and their respective star formation histories are shown in separate panels. b) M32 with a half-light radius of 100 pc is indicated by the red arrow[13]. c) M31's inner stellar halo[14] out to ~30 kpc along the minor axis is indicated by the blue arrow. We represent the star formation history of the inner stellar halo with the 'Brown field' situated 11 kpc along the minor axis[7]. While there are broad similarities, spatial variations exist in the star formation histories of the various inner stellar halo fields[7]. d) The giant stellar stream[15] is indicated by the green solid arrow. e) The inset on the right demonstrates that the metallicity of M32 is much higher than galaxies with similar stellar mass. The solid purple and cyan lines indicate the galaxy metallicity-stellar mass relationship[29,30], while the dashed lines show the 0.18 dex scatter in the relationship. We also indicate the systematic and statistical error in M32's metallicity. The metallicity map of the stellar halo of M31 is reproduced with permission of the authors[11]. Credits of images of M32 and M31: Wikisky.

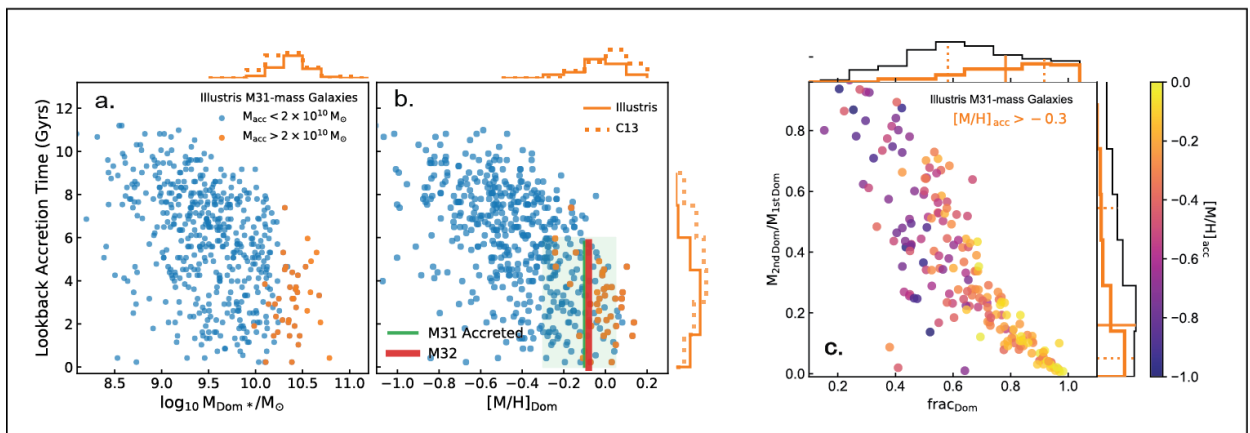

**Figure 2**: **M31's large accreted component and high metallicity constrains its dominant merger to have been a single massive, metal-rich galaxy ($10^{10}<M_{Dom}/M_\odot<5\times10^{10}$, $-0.2<[M/H]_{Dom}<0.2$) accreted in the last ~5 Gyr**. (a-b) The most massive accreted progenitors of M31 analogues are studied using the Illustris simulation[9] (solid orange histogram) and a particle-tagging simulation[10] (dotted orange histograms; labeled C13). We plot the lookback accretion time of the most massive accreted progenitors against their stellar masses and metallicities. The blue and orange symbols show the most massive progenitors of the complete set of M31-mass galaxies in Illustris with $0.73\times10^{12}<M_{virial}/M_\odot<2.21\times10^{12}$, $5\times10^{10}<M_*/M_\odot<2\times10^{11}$ and accreted stellar fraction $f_{acc}<0.5$, showing the full range of possible accretion histories. Orange symbols show the subset of galaxies with large accreted stellar components $M_{*,acc}>2\times10^{10}\,M_\odot$ (See Methods). In panel b, we indicate the estimated metallicity of M31's accreted stellar component and its associated uncertainty derived from the metallicity gradient of its stellar halo by the vertical green line and the green shaded region respectively. We also indicate the constraints on the age of formation of M31's stellar halo derived from the presence of intermediate-age stars by the height of the thick vertical red line. The metallicity and median stellar age[13] of M32 (red) are also indicated. (c) The plot shows the fraction of the second most massive accreted satellite to the most dominant accreted satellite as a function of $frac_{Dom}$ (the fraction of accreted stellar material contributed by the dominant progenitor)[8] for M31-mass galaxies from the Illustris simulations, coloured by the metallicity of the accreted stellar component. The black thin histograms indicate the distribution for all M31-mass galaxies, while the solid orange histogram shows that M31 analogues with a high accreted metallicity ($[M/H]_{acc} > -0.3$; See Methods) are dominated by a single large progenitor. The solid and dashed lines show the median and the 16/84 percentile of the distribution for M31 analogues.

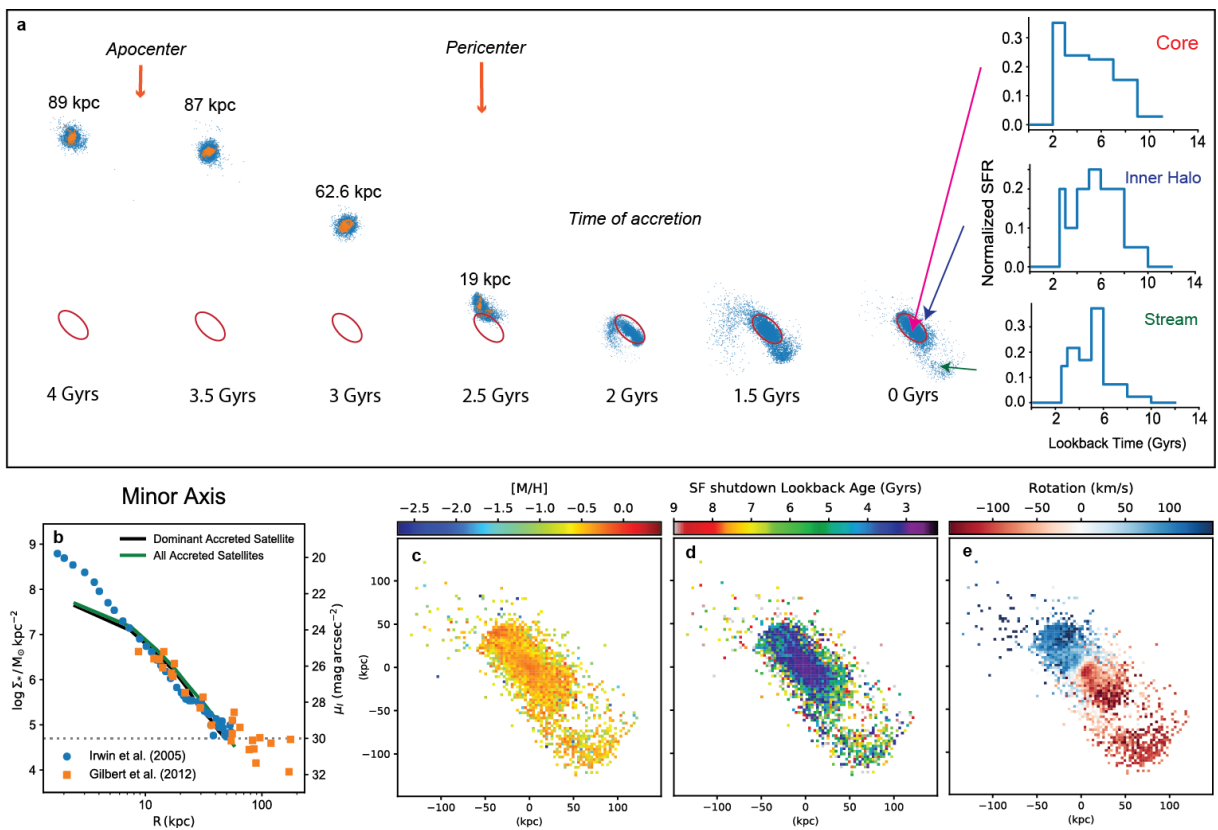

**Figure 3**: **The disruption of the most massive progenitor results in a debris field similar to the stellar halo of M31**. a) We show an example of the tidal disruption of a massive dominant satellite (accompanied by a burst of centrally-concentrated star formation) by an M31 analogue in the Illustris simulation. Blue signifies all the stellar particles of the dominant progenitor while orange signifies those particles that experienced star formation in the last 0.5 Gyr. The red ellipse signifies the extent of the present-day inner stellar halo (40 kpc semi-major axis length). We show the star formation history of the final accreted debris, highlighting the 'core', the 'inner stellar halo' and the 'stream' in order to visually compare with observations presented in Figure 1. b) We also indicate the median stellar mass profile of the final accreted stellar component along the minor axis (green) along with the debris contributed by the massive dominant accreted progenitors (black). We compare this with the M31's minor axis stellar halo profile from the *i*-band surface brightness profile (solid blue circles[19] and orange squares[20]) using $M/L_i=1.8$. The horizontal dotted

line indicates the Illustris mass resolution limit. c-e) show the metallicity, star formation shut-down time and rotation signatures in average velocity from the most massive satellite oriented for a direct comparison with observations in Figure 1. Illustris' resolution does not allow us to resolve compact M32-like bound cores of the disrupted satellite.

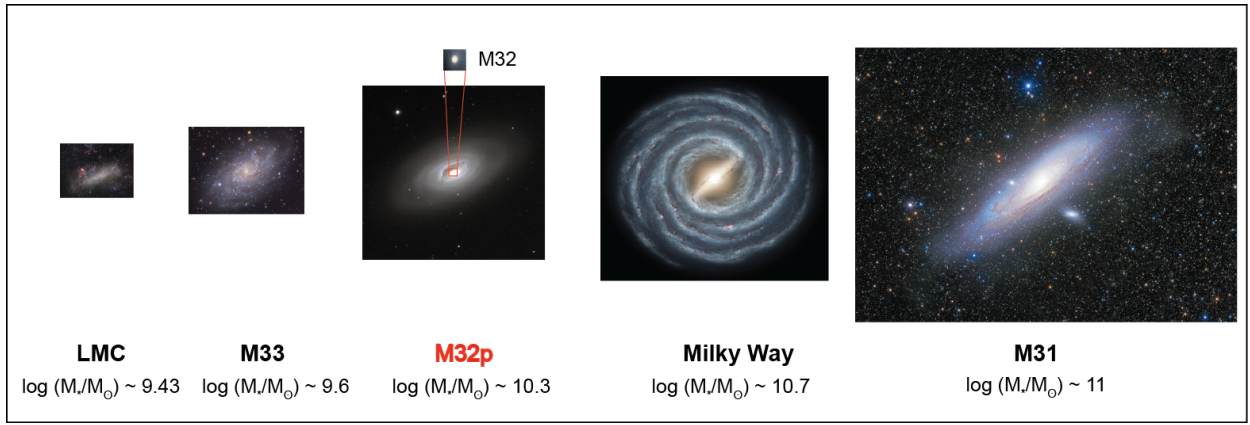

**Figure 4: M32p, the most massive progenitor accreted by M31 was the third largest member of the Local Group.** We compare M32p with the largest present-day members of the Local Group: LMC, M33, MW and M31. We represent M32p using an analogue in the local universe M64. We find a total of 8 M32p analogues out to a volume of ~24 Mpc (Supplementary Fig. 13). The MW is shown using an artist's representation. Credits: LMC, M33 & M31 (Wei-Hao Wang; with permission), MW(NASA/JPL), M64 (NOAO/AURA/NSF).

**Methods:**

**Simulations**

We use two independent cosmological large-scale galaxy formation models, the Illustris hydrodynamical simulations[9,31,32] and particle-tagging simulations[10] (hereafter C13) based on the Munich semi-analytic model[33], to study the global properties and radial distribution of the accreted stellar component of galaxies similar to M31. We discuss the limitations of the models in Supplementary Section 1. For the C13 simulations, we assume the fiducial tagging fraction (5 %) of the most-bound DM particles. We assume a Hubble parameter of 0.72, which affects the masses and the distances of the simulation particles.

In both simulations, we use only the accreted stellar component of central galaxies. Accreted stellar particles are those stellar particles which are formed in subhalos that are not part of the main progenitor branch of the galaxy. The 'dominant' progenitor is the satellite which contributed the most stellar material to the accreted stellar mass of the galaxy. The mass of the satellite is its maximum mass before it is accreted by the main galaxy. We use the median value of the stellar metallicity (all elements above He). The time of accretion of a satellite is when it merges with the main progenitor branch of the galaxy. This corresponds to a time when the satellite cannot be distinguished by the SUBFIND algorithm. Accreted satellites at this stage are usually stripped of most of their stellar material and are within 100 kpc of the host galaxy.

**Selecting M31-mass galaxies**

The strongest constraints on the virial mass of M31 come from the timing argument incorporating the systematic effect from the LMC: $1.33^{+0.39(0.88)}_{-0.33(0.60)} \times 10^{12}$ M$_\odot$ at a 68% (95%) confidence levels[34]. Constraints from SDSS *ugriz* and Spitzer 3.6 µm imaging suggest that M31's stellar mass is 10-15 x 10$^{10}$ M$_\odot$ [35].

We choose M31-mass galaxies from the Illustris and the C13 simulations such that 10.7<log M$_*$<11.3, 11.86<log M$_{DM}$<12.34 and (M$_{acc}$/M*) < 0.5. The last condition on the total accreted stellar fraction ensures that we select only disk-like galaxies. A total of 548 and 680 galaxies satisfy these constraints in Illustris & C13 simulations respectively.

**Constraining M31's total accreted stellar mass**

We constrain M31's total accreted stellar mass from measurements of M31's outer stellar halo. The PAndAS survey estimated the mass of M31's outer stellar halo over a radial range of 27.2 kpc out to 150 kpc, assuming an age of 13(9) Gyr, to be 10.5(8.8) x $10^9$ M$_\odot$[11]. We adopt this radial range to define the outer stellar halo as it both avoids the inner stellar halo and the associated concerns about contributions from an in-situ stellar component and is dominated by the debris of the most dominant accreted progenitor.

The systematic and statistical uncertainties in the radial profiles of the accreted stellar component of our models affect our ability to extrapolate from the mass of the outer stellar halo (R>27 kpc) to the total accreted stellar mass of M31-mass galaxies. Instead, we constrain M31's total accreted stellar component by placing lower and upper limits.

We estimate a lower limit on the mass of M31's total accreted stellar component by using models to determine the ratio of total accreted stellar mass to the accreted stellar mass beyond a projected galactocentric radius of 27 kpc. For Illustris, we find that the total accreted stellar mass is ~0.5 dex larger than that the accreted stellar mass measured beyond 27 kpc. For C13, assuming a fiducial tagging fraction of 5%, we find that the total accreted stellar mass is ~0.65 dex larger than the accreted stellar mass external to 27 kpc. With a smaller tagging fraction, the accreted stellar material is more centrally concentrated. Furthermore, there is more scatter in the ratio of total accreted stellar mass to the stellar mass beyond 27 kpc in the C13 than in Illustris simulations (Supplementary Fig. 1).

In both models, the total accreted stellar mass of a galaxy exceeds the accreted stellar mass outside 27 kpc by at least 0.4 dex. Assuming that the mass of the M31's stellar halo beyond 27 kpc is 8.8x$10^9$ M$_\odot$ [11], we conclude that the total accreted stellar mass of M31 is larger than 2.0x$10^{10}$ M$_\odot$.

In order to set an upper limit to the mass of the accreted stellar component, we require that the accreted stellar component cannot exceed half of the stellar mass of the galaxy ($M_{acc}/M^*<0.5$) to ensure a disk-like morphology for the final galaxy.

**Constraining the metallicity of M31's total accreted stellar component**

The metallicity of the total accreted stellar component is dominated by the metallicity of the most massive accreted progenitor[8]. Galaxies with a large and massive stellar halo have accreted a massive progenitor (for M31, log $M_{Dom,*}$ ~10.3) with a strong metallicity gradient[36]. As this progenitor gets accreted, the metal-rich core of the disrupted satellite sinks to the center[37] leading to strong metallicity gradients in the accreted stellar component of the host galaxy. In such cases, the metallicity of the total accreted stellar component is higher than the metallicity of the outer stellar halo. Using the Illustris simulations, we find that the metallicity of the total accreted stellar component of M31 can be approximated by extrapolating the gradient of the metallicity of the outer stellar halo along the minor axis towards the center of the galaxy[8].

The metallicity of the stellar halo of M31 along the minor axis in the SPLASH[12] survey were derived assuming an age of 10 Gyr and an [α/Fe]=0.0 (Supplementary Fig. 2). However, there is evidence that the outer stellar halo (>60 kpc) is older in age ~10 Gyr, while the inner stellar halo is built up from stars which are considerably younger including intermediate-age stars[7,38], likely from large satellites accreted fairly recently ([α/Fe]~0.0, Age~5 Gyr). This would make the metallicity gradient steeper than previously estimated (green line). The highest metallicity in the outer stellar halo is [Fe/H]~ -0.3. Hence, a robust lower limit for the metallicity of M31's total accreted stellar component is [Fe/H]$_{acc}$> -0.3. Extrapolating the metallicity gradient towards the center of the galaxy[8] suggests that M31's total accreted stellar component has a median metallicity of [Fe/H]$_{acc}$ ~ -0.1 +/- 0.15 dex.

**Choosing M31 analogues**

We choose M31 analogues from our M31-mass galaxies by imposing a lower limit on total accreted stellar mass: log ($M_{acc,*}$)>10.3. We find a total of 39 and 57 galaxies in Illustris & C13 simulations respectively. Of these, 35 and 37 of these galaxies accreted a large satellite (median mass: log $M_{sat}$~10.3, median metallicity: [M/H]~ -0.0) in the last 5 Gyr. We discuss how our selection changes taking M33 into account in Supplementary Section 2.

**Uniqueness of the dominant progenitor**

Although the dominant progenitor contributes most of the mass to the accreted stellar component of M31 analogues, in a few cases the second most massive progenitor can be comparable in mass to the dominant progenitor. We explore this issue by quantifying the fraction of M31 analogues that have had a second massive accretion in the last 5 Gyr above a given mass threshold (Supplementary Fig. 3). For this exercise, we considered only those 'recent' M31 analogues whose dominant accretion was within the last 5 Gyr (90% of M31 analogues). This probability is about 40%(20%) for second accretions with log ($M_*$) > 9.5 for Illustris (C13). This decreases to 10%(7%) for satellites with log ($M_*$)>10. The difference between the two models reflects the steeper slope of the stellar mass function around MW galaxies in the Illustris simulations compared to the C13 model. The second most massive accreted satellite has a mean stellar metallicity of [M/H]~ -0.4 or lower and if present would contribute to accreted features with lower stellar metallicity.

Two pieces of evidence suggests that it is highly probable that M31 is dominated by a single large progenitor.

First, M31's large accreted stellar mass suggests that it is likely dominated by a single large progenitor. We demonstrate this by calculating the fraction of accreted stellar material contributed by the dominant progenitor (frac$_{Dom}$) as a function of accreted stellar mass for M31-mass galaxies in the Illustris and C13 models (Supplementary Fig. 4). M31 analogues are almost all dominated by a single large progenitor. For M31 analogues, frac$_{Dom}$ spans between 0.4 and 1.0 with the mean of the distribution being around 0.8, implying that their stellar halos are built up through the accretion of single large progenitors.

Second, the high metallicity of M31's accreted stellar component also argues that M31's stellar halo is likely dominated by a single large accreted progenitor (Fig. 2c). M31-mass galaxies selected to have a high accreted stellar metallicity component ([M/H]$_{acc}$>-0.3) are dominated by a single large accretion event (high frac$_{Dom}$) and have an average ratio of the stellar mass of the most massive progenitor to the dominant progenitor less than 0.2.

We conclude that it is highly likely that M31 suffered a single large metal-rich accretion event (larger than $10^{10}$ M$_\odot$, [M/H] ~ -0.1) in the last 5 Gyr, which contributed the bulk of the material to its accreted stellar component.

**Properties of the dominant progenitor**

In the Illustris simulations, the majority of the dominant accreted progenitors of M31 analogues are gas-rich (with gas-to-stellar mass ratios spanning a 68% range between 0.4 and 1.3; Supplementary Fig. 5), star-forming, rotating galaxies with pronounced metallicity gradients. These galaxies were star-forming when they were accreted and experienced a peak of star formation at z=1 (Supplementary Fig. 6). The star formation shuts down gradually from the outskirts to the inner parts of the galaxy (see Fig. 3a) around 4 to 6 Gyr ago as they are being accreted by the galaxy. The center of the galaxy tends to be younger and more metal-rich. The centrally-concentrated star formation in these accreted satellites leads to strong star formation history and metallicity gradients (Supplementary Fig. 7).

**Radial profiles of the debris field**

The accreted stellar material contributed by the dominant progenitor determines the bulk properties of the accreted stellar component. We estimate the major and minor axis profiles of the total accreted stellar component for M31 analogues, as well as the accreted stellar profile contributed by the most massive accreted satellite alone (Supplementary Fig. 8). These profiles are calculated in radial bins of 5 kpc using a projected wedge of opening angle of 30 degrees along the major/minor axis. By averaging over the area of our spatial bins, the mass resolution of Illustris allows us to calculate the surface mass density profiles of M31 analogues down to log $\Sigma_*$ ~4.7 M$_\odot$/kpc$^2$ (>3 stellar particles per spatial bin), implying distances out to 65/45 kpc along the major/minor axis. The density profiles are robust to the choice of spatial bin size.

The median major axis profiles of both the total accreted stellar component and the debris field of the dominant progenitor alone are well-fit by exponential profiles beyond 20 kpc. The median minor axis profile of both the accreted stellar profile and the dominant progenitor can be well-approximated by $R^{1/4}$ profiles. The total accreted profiles are very similar to the profile of the dominant accreted satellite alone in the inner parts, with >50% of the accreted mass at a given distance belonging to the dominant accreted satellite within 45 kpc along the minor axis and 70 kpc along the major axis.

There is good agreement between the median minor axis profile of the accreted stellar component of M31 analogues and the observed minor axis surface brightness profile[19,20] of M31's stellar halo (Fig. 3b). This comparison suggests that the minor axis profile beyond a projected distance of 10 kpc (fainter than $\mu_i$~25 mag/arcsec$^2$) appears to be predominantly accreted stellar material. At projected minor-axis distances less than 10 kpc, the surface brightness profiles are dominated by in-situ stellar material.

**Tidal features of the debris field**

Tidal streams are frequently produced in recent large mergers. In Illustris, tidal features are found to survive 3-4 Gyr after the accretion of the galaxy. They extend out to a projected distance of ~100 kpc (Supplementary Fig. 9). They predominantly extend out from the plane of the disk. Furthermore, the present day tidal features frequently do not coincide with orbit of the incoming progenitor[2,39], reflecting the influence of the internal rotation of the progenitor.

**Dynamics of the debris field**

The velocity dispersion of the accreted stellar component is driven by the dispersion of the debris field of the main accreted progenitor. We calculate the projected velocity dispersion of the accreted stellar component of M31 analogues along the minor axis. Out to 50 kpc, we calculate the velocity dispersion in bins of 10x10 kpc$^2$, ensuring a minimum of 5 stellar particles per bin. We calculate the projected velocity dispersion centered at 65 kpc along the minor axis using a bin of 30x30 kpc$^2$. The projected velocity dispersion of the accreted stellar component of M31 analogues is similar to the observational constraints[40,41,42] on the velocity dispersion of the stellar halo from 15 to 65 kpc (Supplementary Fig. 10).

The accreted stellar component of the M31 analogues span a range of rotation velocities, where most have bulk rotation (Supplementary Fig. 11). A small fraction (4/35) of these galaxies are found to possess no significant rotation, while a few galaxies (5/35) exhibit counter-rotating disks. The peak of rotation (>150 km/s) is found at a distance of ~60 kpc along the major axis.

While the dominant progenitor may itself be rotating, the rotation in debris field is caused predominantly by its orbital motion as it is being accreted. Since the rotational velocity of the debris field depends primarily upon the orbit of accretion, it does not constrain well the mass of the progenitor.

The velocity of the outer disk/inner stellar halo of M31 has been well studied[14]. Although a direct comparison of the predicted rotation and the observations is not straightforward, rotation velocities between 150-200 km/s were observed at galactocentric distances of 40-70 kpc along the major axis of M31[14]. These observations are consistent with the wide range in rotation velocities of the accreted stellar component of M31 analogues seen in our models.

**Stellar populations of the debris field**

Strong stellar population gradients are found in the accreted stellar component of M31 analogues. The main driver of these gradients are the stellar population gradients found in the dominant progenitor. The 2D median metallicity maps of the debris from only the dominant progenitor (Supplementary Fig. 12a) show significant differences of nearly a 1 dex in [M/H] from place to place. The bright and prominent tidal features contain preferentially the dominant progenitor's outer low-metallicity stars. Moreover, most of bright tidal features contain a central metal-rich component with a larger metal-poor envelope.

Furthermore, the gradual shutdown of star formation in the dominant progenitor as it is being accreted by the main galaxy leads to variations in the stellar populations across the debris field. In general, accreted stars found at large galactocentric distances are older (>8 Gyr) than stars found at closer radii (<5 Gyr) (Supplementary Fig. 12b). The inner parts of the debris field of the accreted stellar component are significantly younger.

M31's stellar halo has a higher fraction of intermediate-age stellar populations at a projected radius of 11 kpc [7,38] than at larger distances (21 and 35 kpc) [43,44], similar to the accreted debris of our M31 analogues. Given than the tidal features of the dominant progenitor have significantly lower metallicity than its central core, the metallicity of the visible part of the GSS should be lower than the central debris from its progenitor. Furthermore, the shutdown of the bulk of star formation in the Giant Stellar Stream (GSS) at around 5 Gyr is consistent with the cessation of star formation in the outer parts of the dominant progenitor.

**Searching for M32p analogues**

We search for M32p analogues in the local Universe from the S4G survey[45,46]. We define a M32p analogue as a galaxy with stellar mass, $10.2 < \log M_* < 10.6$, and a central surface brightness comparable to M32. Given the I-band central surface brightness profile of M32[23], we require that the surface brightness of the S4G galaxies is greater than 16 mag/arcsec$^2$ in [3.6] band for R<100 pc. The resolution of the S4G survey[45] limits us in probing out to distances of ~24 Mpc. Eight such M32p analogues are found from a total of 115 galaxies in the defined stellar mass range ($10.2 < \log M_* < 10.6$; Supplementary Fig. 13). All eight galaxies exhibit considerable rotation ($V_{rot}$~150-200 km/s). Two are currently interacting with larger primary galaxies (NGC 3034/M82 and NGC 5195/M51b), showing that interactions of M32p-like galaxies with larger primaries at recent times are reasonably common.

There are two known M32-like galaxies out to ~24 Mpc: M32 and NGC4486B. The latter has an effective radius of ~180 pc and a central surface brightness and stellar metallicity comparable to M32[47].

This suggests that the original number of M32p analogues in the volume out to ~24 Mpc was 10. The number density of M32p analogues in this volume is $\log_{10}(N_{M32p}/Mpc^3) = -3.5 +/- 0.316$, while the number density of M32-like galaxies in the same volume is $\log_{10}(N_{M32-like}/Mpc^3) = -4.2 +/- 0.7$ (both number densities incorporate information about the S4G footprint), where the uncertainties quoted are Poisson errors, consistent with the lower limit of the number density calculated from a larger sample of compact ellipticals (see Supplementary Section 3).

The major-merger rate of galaxies of mass log (M*)> 10.0 is approximately 0.1 since z~2 [48]. This suggests that of the original M32p analogue population in the volume extending out to ~24 Mpc, ~1+/-1 of these M32p analogues should have undergone a major-merger since z~2. Hence, the number of M32-like galaxies in the local Universe (~24 Mpc) is consistent with number density of M32p analogues, suggesting that number density of M32p analogues (mid-mass galaxies with a high central surface density) dictates the rarity of M32-like objects in the local Universe.

**Data availability Statement:**

The data that support the plots within this paper and other findings of this study are available from the corresponding author upon reasonable request.

**Additional References:**

# The Andromeda Galaxy's most important merger ~ 2 Gyrs ago as M32's likely progenitor
### (Supplementary Information)

Richard D'Souza[1,2], Eric F. Bell[1]

[1]Department of Astronomy, University of Michigan, 311 West Hall, 1085 S. University, Ann Arbor, MI 48109-1107, USA

[2]Vatican Observatory, Specola Vaticana V-00120, Vatican City State.

# 1. Advantages and Limitations of the models

In this work, we use two independent cosmological large-scale galaxy formation models, the Illustris hydrodynamical simulations[1] and particle-tagging simulations[2] (hereafter C13), to study the global properties and radial distribution of the accreted stellar component of galaxies similar to M31. In particular, we use these models to demonstrate that the mass and metallicity of M31's total accreted stellar component constrains the mass of its most dominant merger.

Hence, we are interested only in the bulk properties (mass and metallicity) of M31's stellar halo. This choice of metrics is less subject to systematic error, since they are independent of the exact positions, orbits and motion of the accreted stars. Consequently, while we are less concerned with our simulations reproducing the exact phase-space information of M31's stellar halo, we cannot constrain the exact orbit of M31's most massive merger event.

The two simulations are particularly suitable to study the bulk properties of the accreted stellar component of M31-mass galaxies for the following reasons. a) Due to their large volume, they encode a diversity of accretion histories. b) They represent the accretion histories of M31-mass galaxies reasonably well: both simulations have approximately the right halo occupation of accreted satellites of M31-galaxies enforced through the galaxy stellar mass function (9<log M* <10) and the cosmic star formation history. c) They reproduce the stellar-mass metallicity relationship of galaxies fairly well[1,3,4]. d) They have enough resolution to resolve the general properties of the most significant progenitors of M31-mass galaxies.

On the other hand, the radial profile of the accreted stellar component depends upon the tidal disruption of satellites and is still highly model dependent: a) The disruption of accreted satellites depends upon the satellite galaxies having the right sizes and shapes (correct binding energies). b) The distribution of the debris depends upon the model getting the right potential of the galaxy. The physical extent of galaxies in the Illustris simulations can be a factor of a few larger than observed for $M_* < 10^{10.7}$ M$_\odot$ [5], affecting the spatial distribution of the accreted stellar debris. The C13 simulation is limited in its ability to reproduce the 3D distribution of the accreted stellar debris, as its galaxies assume the shape of the inner part of their dark matter halos and do not account for the potential of galactic disks. Furthermore, the mass resolution of the simulations (~ $10^6$ M$_\odot$) becomes a limiting factor in studying the phase-space distribution of the debris of the most massive progenitor. These simulations also cannot resolve M32-like compact cores in the progenitors, or their remnants in the final stellar halos.

Since the radial profiles of the accreted stellar component are subject to substantially larger systematic errors than the total accreted stellar mass and metallicity, the use of the outer stellar halo mass alone to choose M31 analogues would yield a sample more subject to systematic differences from model to model than a set selected on the more robust total accreted stellar mass alone. Accordingly, we choose to isolate M31 analogues to have a constrained range of total accreted stellar masses.

2. **Considering the role of M33 in the choosing M31 analogues**

Imposing that these M31 analogues have also a current satellite of the size of M33 (log M* ~ 9.6) decreases that number to 8 and 13 galaxies in the two simulations respectively. All of these galaxies accreted a large satellite (median stellar mass: log $M_{sat}$~10.36, median metallicity [M/H]~ -0.0) in the last 5 Gyr. We conclude that it is possible for M31 to have both a large satellite like M33 as well as have accreted a large progenitor (log M*~10.3) in the last 5 Gyr. Furthermore, it does not change any of the findings of this work, but only decreases our number statistics.

3  **Compact Ellipticals**

M32 is classified as a compact elliptical galaxy. This class of galaxies are quite rare, with only 200 objects presently known[6-8]. They are small, compact (100 pc < $R_{eff}$ < 500pc), high stellar density, non-star-forming galaxies with low stellar masses ($10^8$ < M*/ M$_\odot$ < $10^{10}$). In this stellar mass range, they set themselves apart distinctly from dwarf ellipticals as well as ultra-compact dwarfs. While most of these galaxies are found in clusters or in the haloes of their host galaxies, a few compact ellipticals have also been found in the field[8,9,10].

In the literature, the term compact ellipticals is also applied to a broader set of galaxies, with a looser definition of compactness, with some researchers extending the effective radius range $R_{eff}$ < 1.5 kpc [11,12]. Some of these galaxies may be relic galaxies of high-z, massive compact galaxies[13].

The discovery of isolated compact ellipticals presents a challenge to the stripping mechanism. Although some of these isolated compact ellipticals are compatible with being galaxies that have been ejected from their environments where they were originally stripped[8], it is difficult to totally rule out other mechanisms.

We estimate the number density of compact ellipticals discovered so far. For this we use a large sample (~195) of compact elliptical galaxies ($R_{eff}$ < 0.5 kpc) found in all types of environments including in the field, which were discovered by mining large-survey data[8]. We calculate an approximate number density of these compact ellipticals between z=0.03 and z=0.05, where the completeness of the methodology of detection is relatively high. Our estimated lower limit of the number density, incorporating information about the SDSS survey area, is $\log_{10}(N/Mpc^3)$ = -4.6 +/- 0.1. This lower limit is compatible with the number of M32-like objects calculated locally in the volume out to ~24 Mpc (see Methods section).

Moreover, this number density of compact ellipticals is also consistent with number density of M32p analogues calculated earlier, with 1/10th of them having undergone a merger with a larger galaxy since z~2. This implies that the stripping scenario could explain the number density of compact ellipticals found in the literature, suggesting that it is the dominant mechanism for the creation of compact ellipticals. Similar conclusions were also derived from the fraction of compact ellipticals found in the vicinity of a larger host[8].

4  **Dynamical models of the Giant Stellar Stream**

The Giant Stellar Stream (GSS), owing to its high metallicity, has a relatively high *a priori* probability of originating from M32p (See Methods). There is a roughly 10-30% *a priori* possibility that the GSS could originate from other recently accreted lower-mass progenitors (depending on the mass of the required progenitor, 9.5< log M*<10.0; See Supplementary Fig.

3). In practice, the probability of associating the GSS and M32p should be significantly larger, owing to the similarity of the metallicities and star formation histories of the GSS and parts of the inner stellar halo (Fig.1).

If indeed the Giant Stellar Stream (GSS) is associated with M32p, it offers a unique opportunity to constrain M32p's orbit. This motivates a survey of existing frameworks for dynamically modeling and constraining the orbit of the progenitor of the GSS.

The availability of line-of-sight distances[14,15] and radial velocity measurements[16] of the GSS have motivated several attempts to dynamically model the accretion of a dwarf galaxy in order to constrain its properties. However, since the uncertainties in the TRGB distances[15] allow for considerable range of orbits[17], it remains extremely challenging to constrain the properties of the progenitor solely on observations of the GSS alone. In order to make further progress, models resort to additional observational constraints. The first category of models uses the response of M31's disk to the accreted satellite, while the second category constrains the orbit by identifying possible forward debris associated with the GSS.

**Disk models:** These set of methods assume that the progenitor of the GSS was the most massive accreted satellite to have perturbed M31's disk. a) The models of Mori & Rich use the thickness of M31's disk to constrain the mass of the stream's progenitor to $M_* \leq 5 \times 10^9$ $M_\odot$[18]. Yet this low mass constraint is inconsistent with the implications of M31's large stellar halo, indicating that the physics of disk thickening due to satellite interaction is complex and poorly understood[19,20]. B) Hammer et al. constrain the mass of the accreted progenitor using the size of M31's HI disk[21,22]. Their models attempt to reproduce the thick disk, the 10 kpc star-forming ring, the bulge, the GSS as well as the stellar halo through the merger of a single and massive gas-rich progenitor[21]. They demonstrate that the steep age-velocity dispersion of M31's disk puts stringent constraints on the timing of its most massive merger; they suggest a fairly recent massive merger (1.8-3 Gyr ago, 1:4) [22] in agreement with this work.

**Other Tidal Debris models:** These methods work by identifying the forward tidal debris of the GSS. The predicted debris field of the progenitor in these models is similar in morphology to M31's North-Eastern (NE) and Western (W) shelves. Moreover, their stellar populations are fairly similar to the GSS allowing one to empirically associate them with the debris of the stream's progenitor[23-29]. With a suitable model of M31's potential[30], the incoming orbit is determined taking into account the width of the GSS caused by the internal motion of the progenitor[23]. The models of the accreted progenitor range from a simple Plummer sphere[24] to a rotating disk-bulge-halo configuration ($V_{rot} \sim 60$ km/s) to account for the azimuthal distribution of debris in the GSS[27-29]. By comparing the surface brightness between the W shelf (the forward debris of the progenitor) and the GSS (the trailing stream of the progenitor), they constrain the progenitor's mass and orbit[26]. While the models differ in the details, they constrain the mass of the progenitor at its last pericentric passage ($t_{acc} \sim 0.7$ Gyrs) as $\sim 5 \times 10^9$ $M_\odot$, whose remnant should be in the NE shelf[26].

**Reconciling the two models:** Not only does the major-merger model differ from dynamical models of the debris of the stream differ with regard to the *mass* of their progenitors, but there are also significant differences in their internal motions (e.g. rotation) which dramatically affect the distribution of the debris field and the inferred *orbit*. While it is difficult to compare the two sets of models, intuition may be drawn from the GSS. Both sets of models reproduce the complex features of the GSS with very different stellar masses and rotations of their progenitors[22,26]. This suggests that it is difficult to constrain the stellar mass of the progenitor and rotation simultaneously from observations of the GSS alone. Constraints based solely on the velocity dispersion of the GSS may be misleading as the stream could be built up from the outer low velocity parts of a massive progenitor (see Fig. 3). Furthermore, progenitors as large as M32p with a significant internal rotation (as large as 150-200 km/s) can dramatically alter the distribution of tidal debris in the NE and W shelves, affecting the orbit and mass of the progenitor inferred by comparison of only the GSS and the W shelf.

Indeed, the metallicity gradients present in the GSS suggest an original progenitor which a much larger stellar mass. Intriguingly, the tidal-debris models dynamically admit much larger progenitors of the GSS (log $M_{Total} \sim 10.3$ at 95% confidence limit, see Fig. 5 of Fardal et al.[26]), though the amount of stellar material present in the GSS and the W shelf do not support a dispersion-supported progenitor of the same stellar mass[26].

The observed positions and velocity caustics of the NE and W shelves[26] frame the range of possible orbits of the progenitor. M32's position and radial velocity are consistent with the empirical limits derived from the two shelves. Furthermore, if the NE and W shelves are part of the trailing and forward debris of the progenitor respectively, then M32's proper motion should be consistent with this motion.

**Next generation of models:** The findings of this work suggest two significant priors for the next generation of dynamical models of the GSS. First, M31's large stellar halo should inform a prior favoring much a larger stellar mass for the progenitor with significant rotation. Second, the presence of a compact dense galaxy like M32 in the field suggests a suitable prior on the central density of progenitor. Current models suggest that progenitors with extremely dense central nuclei survive much longer in a tidal field[26], leaving behind a compact core which does not disrupt so easily, paralleling the analogous formation of ultra-compact dwarf galaxies[31].

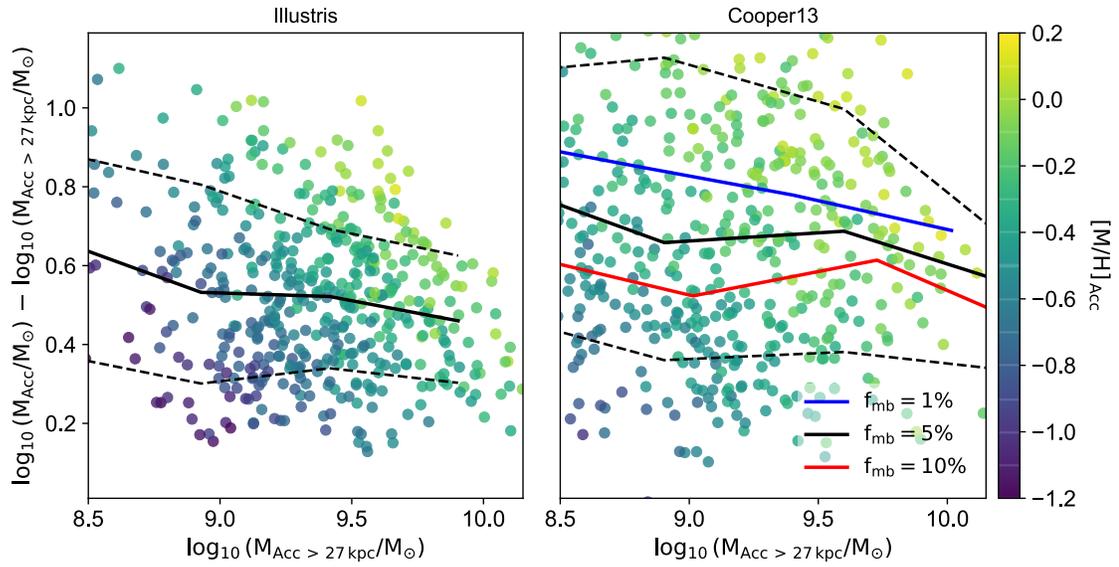

**Supplementary Figure 1: Estimating the total accreted stellar mass of a galaxy from aperture measurements of the outer stellar halo.** The plot shows the difference between the total accreted stellar and the accreted stellar mass measured beyond a projected radius of 27 kpc for M31-mass galaxies in the Illustris (left panel) and the C13 (right panel) simulations. In the left panel, the solid black line shows the median of the distribution while the dashed lines show the 16th and 86th percentile of the distribution. In the right panel, the solid and dashed black lines show the median and the 16th/86th percentile of the distribution in the C13 simulations for the fiducial tagging fraction (5%) of the most bound DM particles. Additionally, the blue and red lines show the medians of the distribution in the C13 simulations when 1 and 10% tagging fractions were employed. On both panels, the points are colored by the metallicity of the total accreted stellar component of the galaxies.

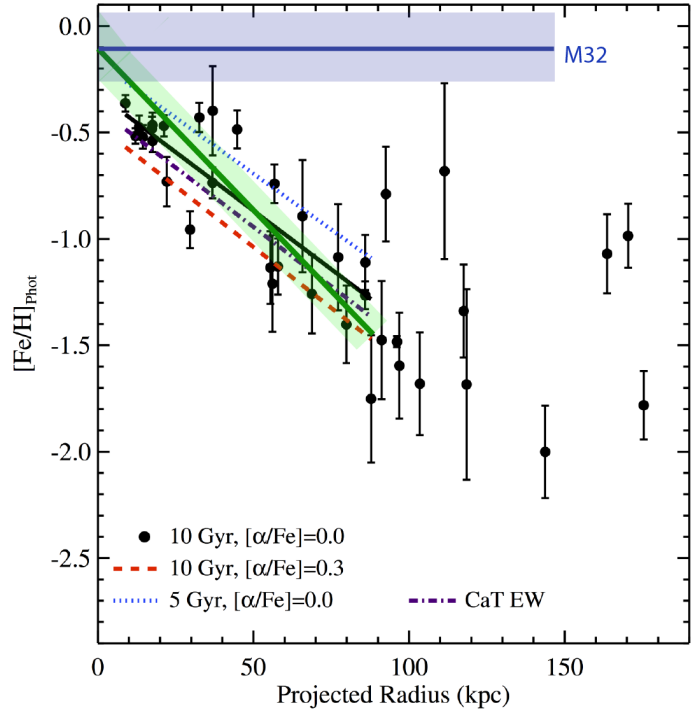

**Supplementary Figure 2: Estimating the metallicity of the total accreted stellar component of M31 using the minor axis metallicity gradient of its stellar halo along the minor axis**. This plot has been adapted from the SPLASH survey and has been reproduced with permission from the authors[32]. The black line is the photometric metallicity gradient assuming a stellar population with an age of 10 Gyr and [α/Fe]=0.0. The red and blue lines represent variations in the photometric metallicity gradient by allowing for a higher [α/Fe] (~0.3) and a younger stellar population (age ~ 5 Gyr) respectively. The dotted-dashed line represents the spectroscopic metallicity gradient derived from the CaT lines. Assuming that the outer stellar halo contains older populations, while the inner stellar halo is made up of intermediate-age populations[33], the green line and the accompanying shaded region represents our best-estimate along with the confidence limits (statistical errors) of the metallicity of M31's outer stellar halo. The intercept of the green line at R=0 gives the estimated metallicity of M31's total accreted stellar component. M32's metallicity and its uncertainty are also indicated by the horizontal blue line and its accompanying shaded region.

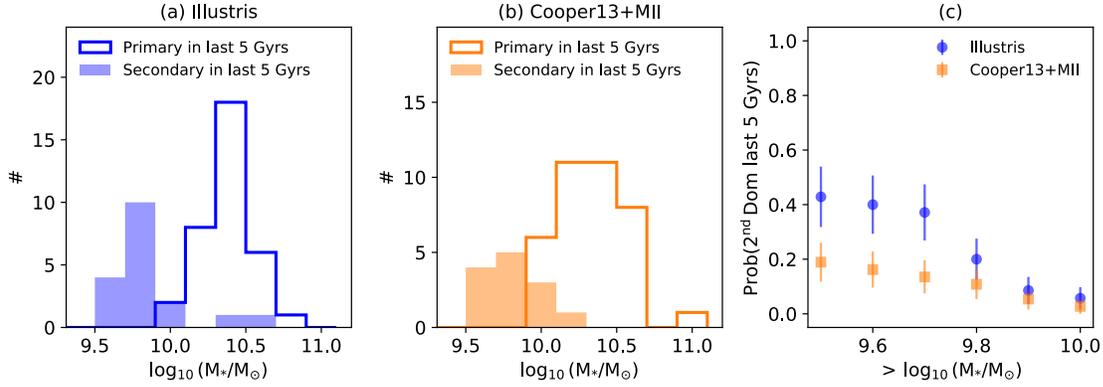

**Supplementary Figure 3**: **The properties of the second most-massive satellite accreted in the last 5 Gyr by M31 analogues.** a) A histogram of the most massive (primary, blue-outline) and the second most massive (secondary, blue-filled) satellites recently accreted by M31 analogues in the Illustris hydrodynamical simulation. b) A similar histogram of the primary (orange-outline) and secondary (orange-filled) recently accreted satellites in the C13 simulation. c) The estimated probability that the secondary accreted progenitor could have contributed to the dominant features of the stellar halo of M31 analogues as a function of a threshold mass in the Illustris (blue points) and C13 (orange points) models.

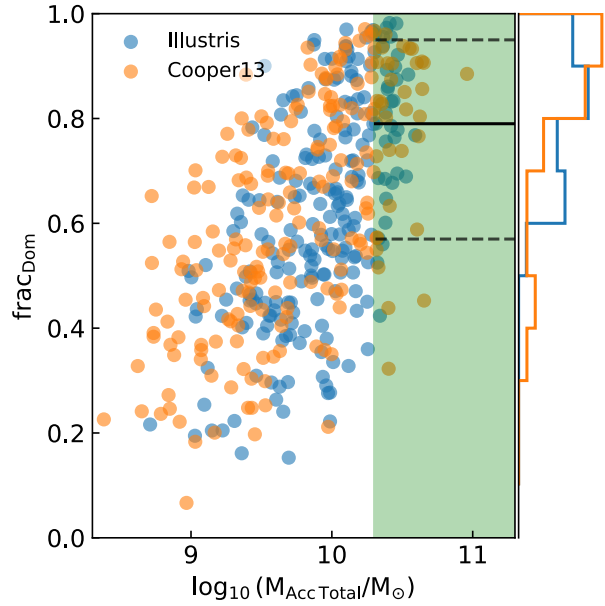

**Supplementary Figure 4: The contribution of the dominant accreted progenitor to the accreted stellar halo.** We plot the fraction of stellar mass contributed by the dominant progenitor to the total accreted stellar component (frac$_{Dom}$) for M31-mass galaxies for the Illustris (blue) and C13 (orange) simulations. The green shaded region indicates M31 analogues (log M$_{acc}$ > 10.3). The solid and dashed black horizontal lines show the median and 10/90 percentile of the distribution of frac$_{Dom}$ for Illustris M31 analogues.

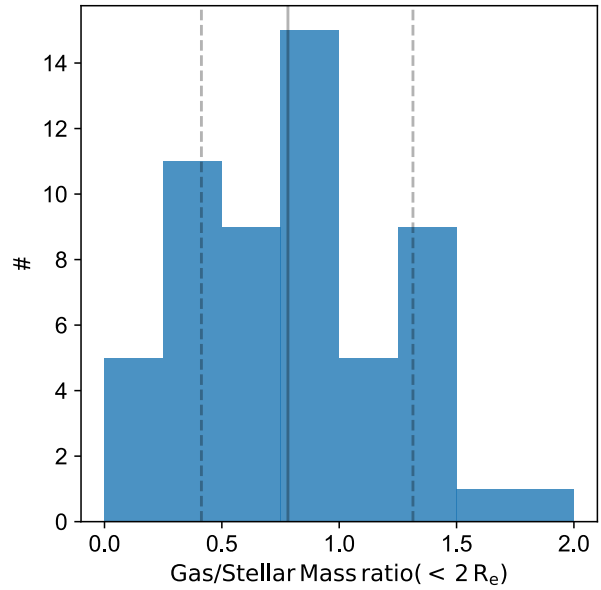

**Supplementary Fig. 5: The gas-to-stellar mass ratios of the dominant accreted progenitor of Illustris M31 analogues.** The gas-to-stellar mass ratios were estimated within a sphere enclosing twice the effective radius when the galaxy had its maximum stellar mass. The solid vertical line indicates the median of the distribution, while the dashed vertical lines indicates the 16 and 84 percentiles of the distribution.

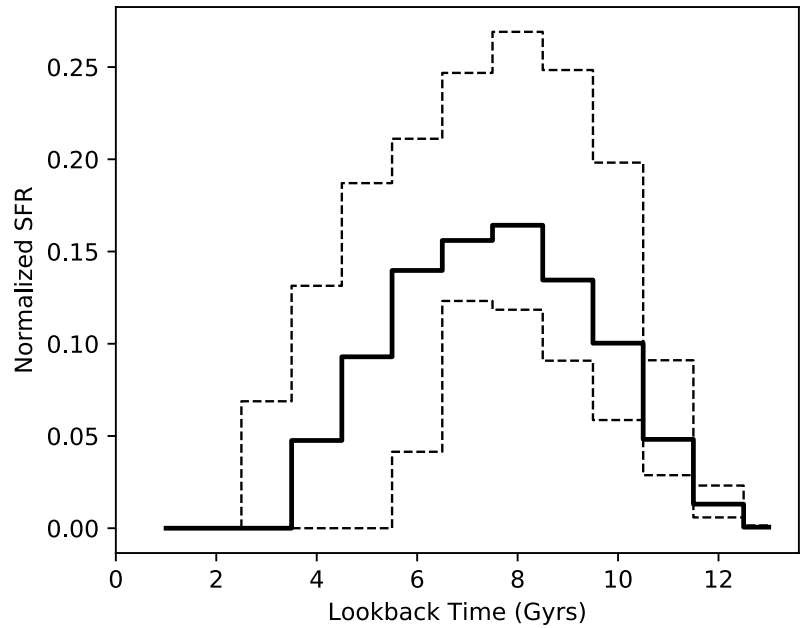

**Supplementary Figure 6: The star-formation history of the dominant progenitors of Illustris M31 analogues.** The solid black line shows the median star-formation rate of the dominant progenitors of M31 analogues as a function of lookback time, while the dashed lines show the 10 and the 90% percentile of the distribution of star-formation rates.

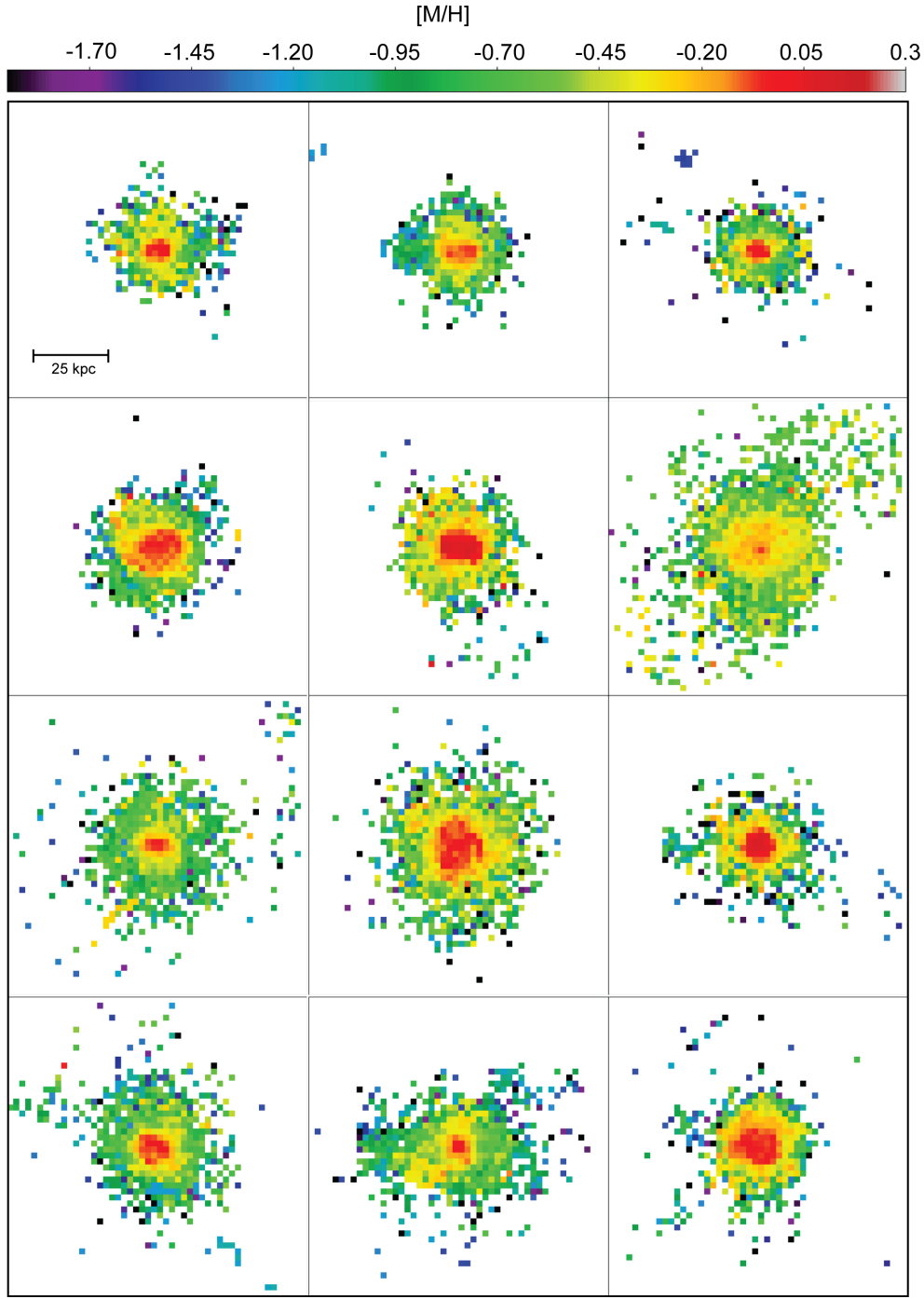

**Supplementary Figure 7: Metallicity gradients in the dominant accreted progenitors of M31 analogues.** Face-on median metallicity maps of selected dominant accreted satellites which are later accreted by the Illustris M31 analogues.

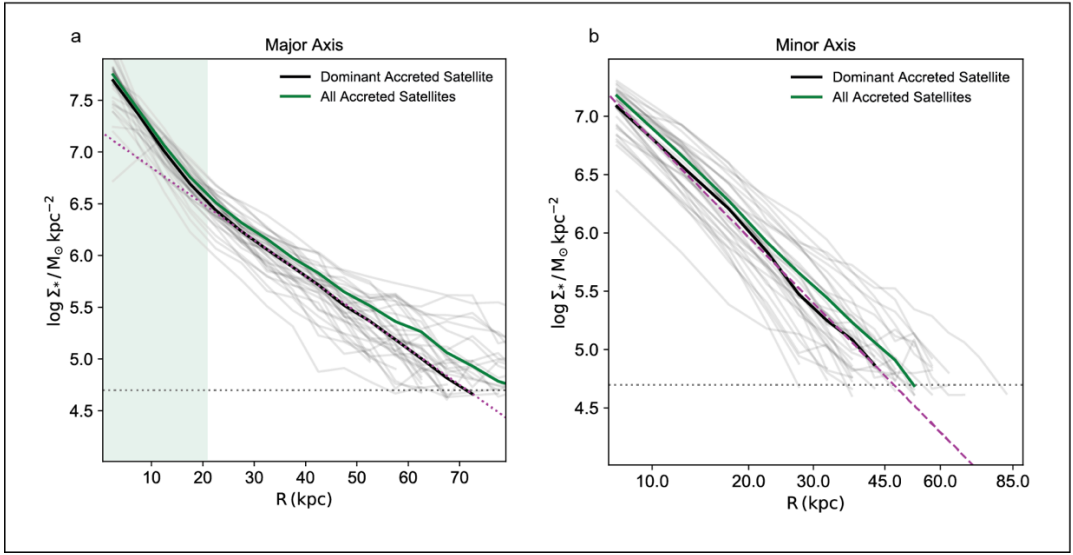

**Supplementary Figure 8: The contribution of the dominant accreted progenitor to the radial profiles of the accreted stellar component in the Illustris simulations.** The left and right panels show the median radial profile of the accreted stellar component along the major and minor axis respectively. The solid green line shows the median profile of the total accreted stellar component while the black solid line indicates the median contribution of the dominant accreted progenitor. The light black lines show the individual radial profiles of the stellar debris of the dominant accreted progenitors. To guide the eye, the dashed magenta line in the left panel is an exponential fit to the median radial profile of the most massive accreted satellite beyond 20 kpc, while the dashed magenta line is a de Vaucouleurs fit to the median radial profile of the most massive accreted satellite (the x-axis is linear in $R^{1/4}$). We also indicate the mass resolution limit of the Illustris simulations by the horizontal dotted line.

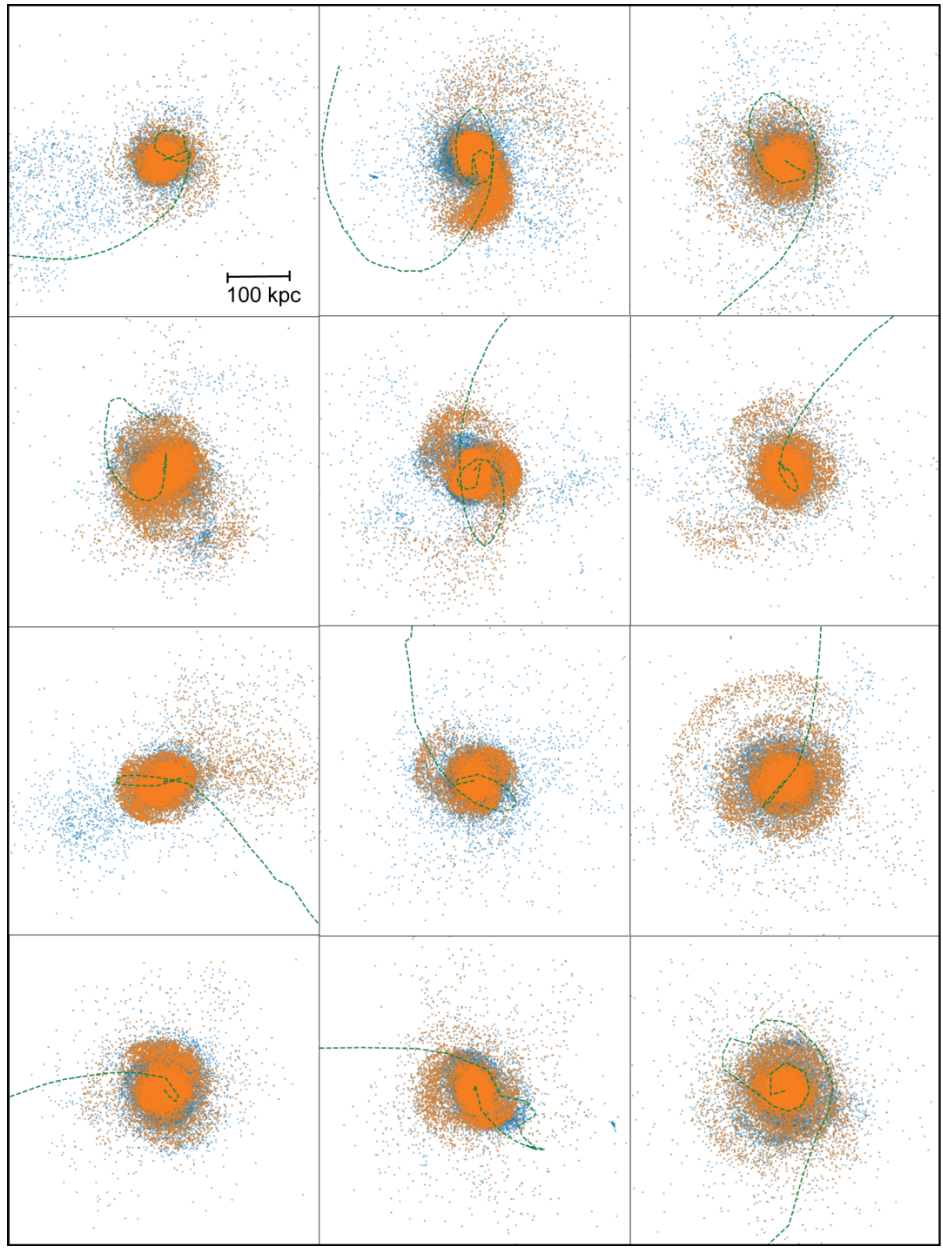

**Supplementary Figure 9: Tidal features in Illustris M31 analogues.** A face-on view of the debris of the accreted stellar component of M31 analogues. The blue dots represent all accreted stellar particles, while the orange dots represent stellar particles which belong to the dominant progenitor. The green line traces the projected path of the dominant progenitor as it is accreted by the M31 analogue.

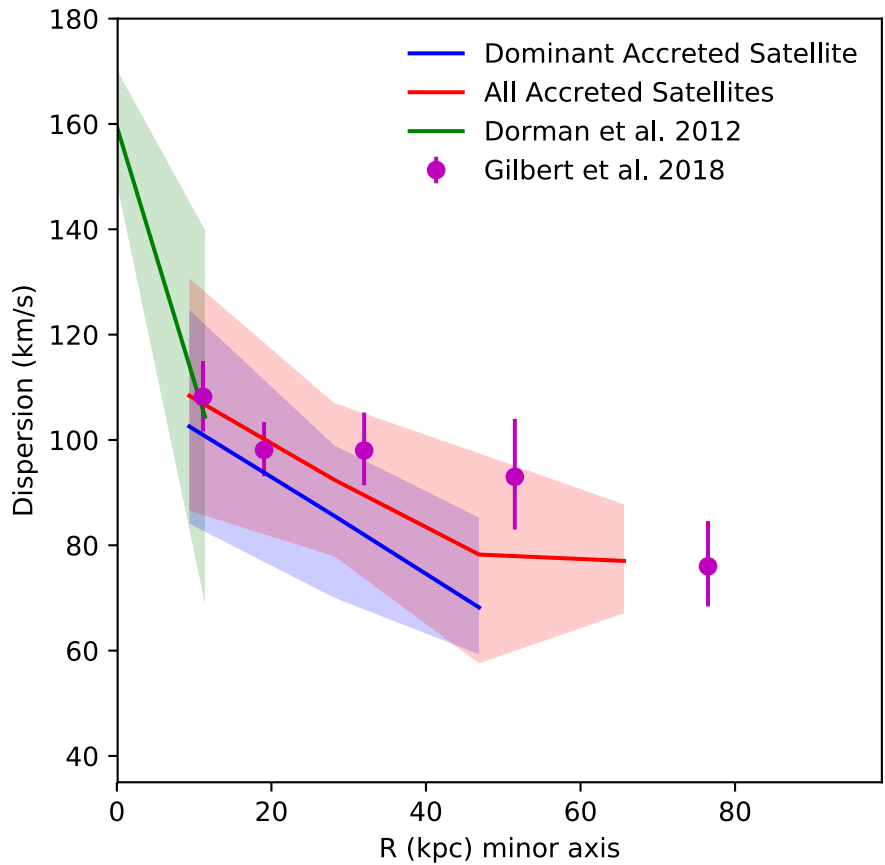

**Supplementary Figure 10: Velocity dispersion profile along the minor axis**. The red line shows the median velocity dispersion as a function of radius along the minor axis of the total accreted stellar component of M31 analogues in the Illustris simulations. The blue line shows the median velocity dispersion profile contributed by the dominant accreted progenitor. The red and blue shaded regions show the 10 and 90 percentiles of the distributions. We also indicate the velocity dispersion of M31's inner and outer stellar halo with the green line and the magenta points respectively.

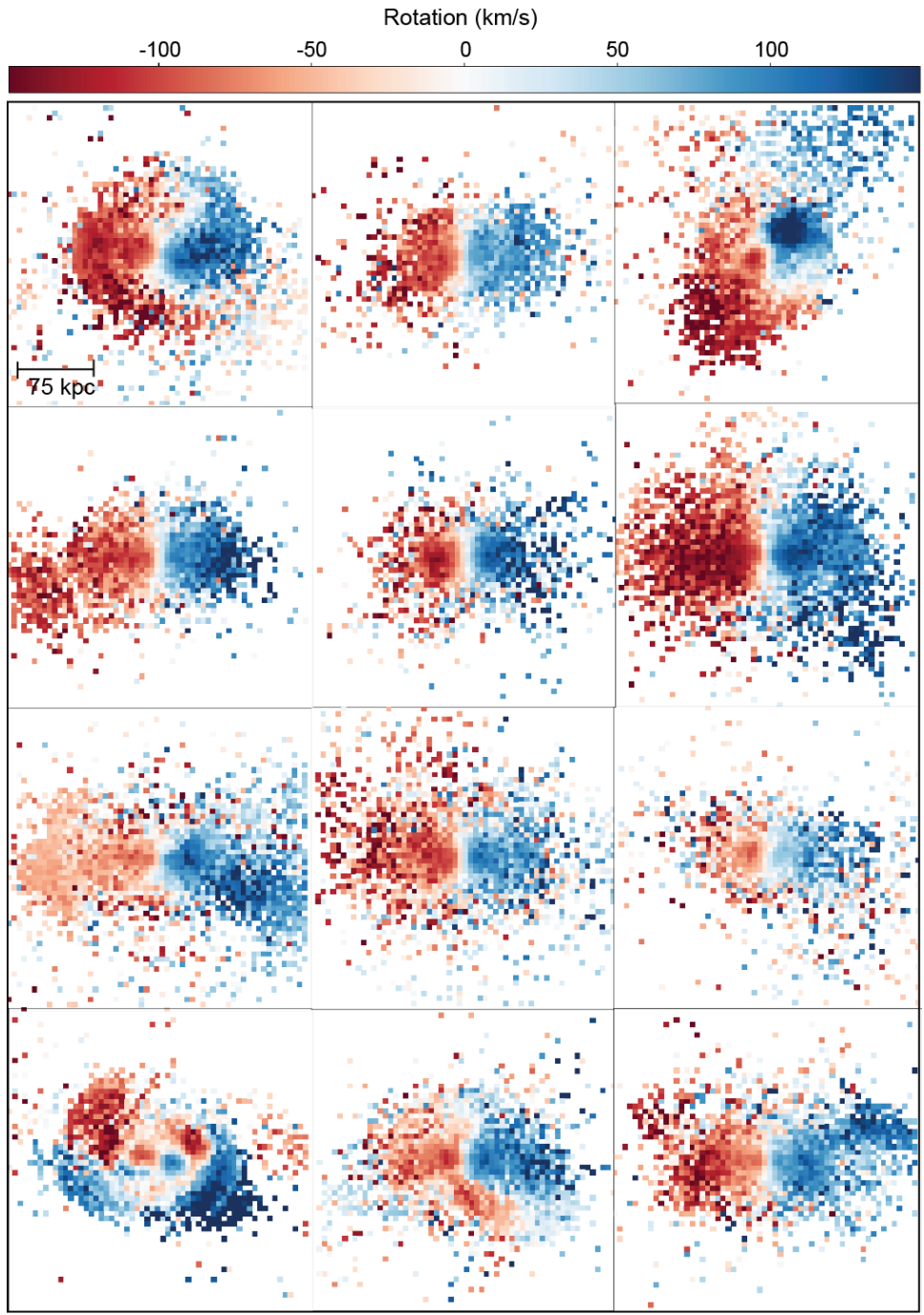

**Supplementary Figure 11**: **2D rotation maps of Illustris M31 analgoues.** We show an edge-on view of the accreted stellar component of selected M31 analogues (Illustris). The galaxies were tilted to an inclination angle of 78 degrees, in common with M31.

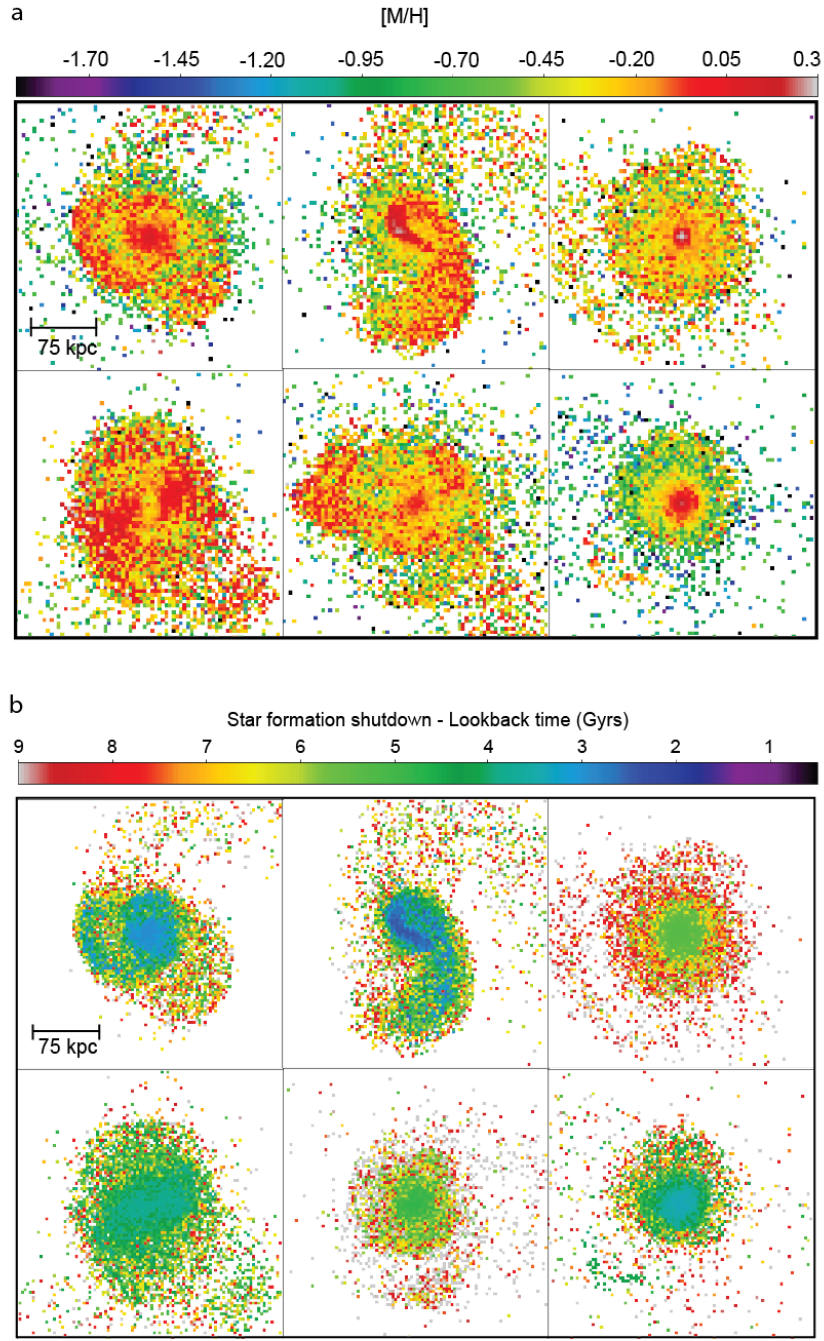

**Supplementary Figure 12: Face on 2D maps of the debris field of the dominant accreted progenitor for a selection of Illustris M31 analogues.** Panel a shows the median metallicity while panel b shows the median lookback time for the shutdown of star formation in the debris field contributed by the dominant accreted progenitor.

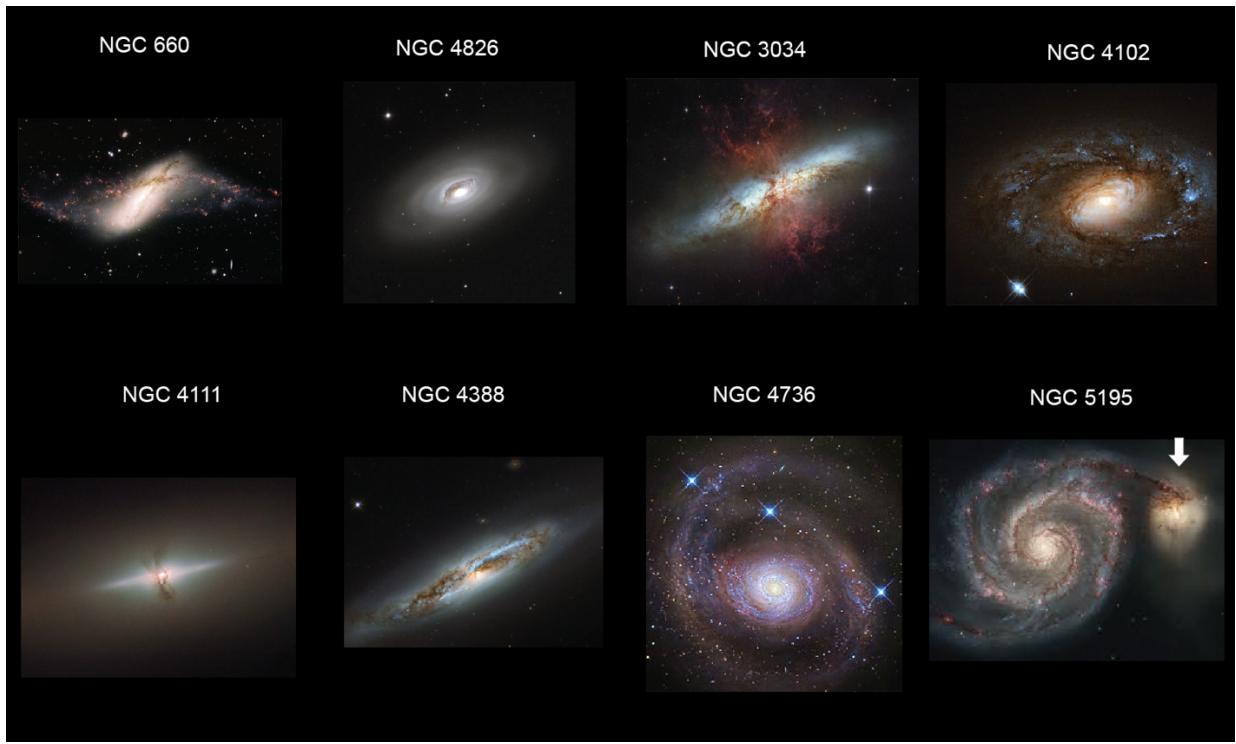

**Supplementary Figure 13: M32p analogues selected from the S4G survey**. The images are taken from the public domain. Credits: NGC 660 (Gemini Observatory), NGC 4826 (NOAO/AURA/NSF), NGC 3034, NGC 4012, NGC 4111, NGC 4388 & NGC 5195 (ESA/Hubble & NASA), NGC 4736 (Jay Gabany).